# Brain Principles Programming

Evgenii Vityaev[1,2][0000-0002-5651-6781], Anton Kolonin[2][0000-0003-4180-2870], Andrey Kurpatov[3], Artem Molchanov[3][0000-0003-0197-8871]

[1] Sobolev institute of mathematics, Koptuga 4, Novosibirsk, Russia
[2] Novosibirsk State University, Pirogova 2, Novosibirsk, Russia
[3] Sberbank of Russia, Neuroscience Lab, Moscow, Russia
vityaev@math.nsc.ru

**Abstract.** The monograph "Strong Artificial Intelligence. On the Approaches to Superintelligence", referenced by this paper, provides a cross-disciplinary review of Artificial General Intelligence (AGI). As an anthropomorphic direction of research, it considers Brain Principles Programming (BPP) – the formalization of universal mechanisms (principles) of the brain's work with information, which are implemented at all levels of the organization of nervous tissue. This monograph provides a formalization of these principles in terms of the category theory. However, this formalization is not enough to develop algorithms for working with information. In this paper, for the description and modeling of BPP, it is proposed to apply mathematical models and algorithms developed by us earlier that modeling cognitive functions, which are based on well-known physiological, psychological and other natural science theories. The paper uses mathematical models and algorithms of the following theories: P.K.Anokhin's Theory of Functional Brain Systems, Eleonor Rosh's prototypical categorization theory, Bob Rehter's theory of causal models and "natural" classification. As a result, the formalization of the BPP is obtained and computer examples are given that demonstrate the algorithms operation.

**Keywords:** Brain principles, categorization, category theory, formal concepts.

## 1 Introduction

In the monograph "Strong Artificial Intelligence. On the approaches to superintelligence" [1] the first cross-disciplinary guide on Artificial General Intelligence (AGI) is given: "General artificial intelligence is the next step in the development of AI, not necessarily endowed with self-awareness, but, unlike modern neural networks, capable of coping with a wide range of tasks in different conditions." As an anthropomorphic direction of research, it considers Brain Principles Programming (BPP) – the formalization of universal mechanisms (principles) of the brain's work with information, which are implemented at all levels of the organization of nervous tissue. The book provides a formalization of these principles in terms of category theory. However, algorithms for working with information do not follow from this formalization.

In this paper, we apply mathematical models and algorithms developed by us earlier that modeling cognitive functions and based on the well-known physiological, psychological and natural science theories, to model Brain Principles Programming. We will rely on the following theories: P.K.Anokhin's Theory of Functional Brain Systems [2-3], Eleonor Rosh's prototypical categorization theory [4-5], Bob Rehter's theory of causal models [6-7] and works on "natural" classification [8].

# Part I. Basic theories and formal models

## 1.1 Basic elements of perception and the world

**1.1.1. "Natural" classification.** The first philosophical analysis of the "natural" classification belongs to J.S. Mill [9]: “"artificial" classifications differ from "natural" in that they can be based on ... features, so that different classes differ in that they include objects with different meanings of these features. But if we consider the classes of "animals" or "plants", then they differ in such a large (potentially infinite) number of properties that they cannot be enumerated. And all these properties will be based on statements confirming this difference”.

J.S. Mill's analysis has been confirmed by naturalists. L. Rutkovsky writes about the similarity of properties in "natural" classes [10]: "The more essential features of the compared objects are similar, the more likely their similarity is in other features." Smirnov E.S. [11]: "The taxonomic problem lies in the "indication": from an infinitely many number of features, we need to move to a limited number of them, which would replace all other features". From studies on "natural" classification, it follows that features in "natural" classes are strongly correlated, since a potentially infinite number of features are almost uniquely determined follow L. Rutkovsky from the indicator features. A formal model of "natural" classification is given in [6].

**1.1.2. "Natural" concepts and prototypical theory of categorization.** The high correlation of features for "natural" classes has been confirmed in cognitive studies. In the works of Eleanor Rosch [4-5], the following principle of categorization of "natural" categories was formulated, confirming the statements of J.S. Mill and naturalists: "Perceived World Structure: … perceived world – is not an unstructured total set of equiprobable co-occurring attributes. Rather, the material objects of the world are perceived to possess ... ***high correlational structure*** (emphasis added by EV). … In short, combinations of what we perceive as the attributes of real objects do not occur uniformly. Some pairs, triples, etc., are quite probable, appearing in combination sometimes with one, sometimes another attribute; others are rare; others logically cannot or empirically do not occur». Therefore, directly perceived objects (so called basic objects) are information–rich bundles of observable and functional properties that form a natural discontinuity that creates categorization. These bundles form "prototypes" of class objects. Further Eleanor Rosch's theory of "natural" concepts was called the prototypical theory of concepts (prototype theory).

**1.1.3. The theory of causal models.** Studies have shown that people's knowledge of categories is not reduced to the set of features, but includes a rich set of causal relationships between these features. In some experiments [7] it was shown that a feature is more important if it is more strongly included in the causal network of interrelations of features. Considering these studies, Bob Rehder put forward the theory of causal models (causal-model theory), according to which the relation of an object to a category is no longer based on a set of features and proximity by features, but on the basis of similarity of the generative causal mechanism [7]. In [8] Bayesian networks were used to represent causal knowledge. However, they cannot model cyclic causal relationships because Bayesian networks do not support cycles. The probabilistic formal concepts, that developed earlier and presented in the supplement of work [12], model cyclic causal relationships by fixed points of causal relationships predictions.

### 1.2 Probabilistic formal concepts

We assume that "natural" classification and "natural" concepts are described by the same formalism. From our point of view, the information processes of the brain were tuned in the process of evolution to extract a highly correlated structure of features of "natural" objects by forming "natural" concepts of these objects. Causal relationships in the perception of "natural" objects close on themselves forming a certain "resonance". At the same time, "resonance" occurs if and only if these causal relationships reflect some holistic "natural" object. The resulting cycles of inferences on causal relationships are mathematically described by "fixed points" of mutually interconnected features, which gives an "image" or "prototype" of objects class. Therefore, the brain perceives a "natural" object not as a set of features, but as a "resonating" system of causal relationships that forms a causal model.

The formalization of cyclic causal relationships in the form of probabilistic formal concepts is given in the supplement of the work [12]. There is also a definition of causality in the form of the Most Specific Causal Relationships (MSCR) that solve the problem of statistical ambiguity (see references in the supplement). There is also an example of the probabilistic formal concepts discovery.

Probabilistic formal concepts can be used to context (latin contextus – cohesion, coherence, connection) detection as systems of interrelated concepts in a certain situation, discourse, a certain point of view, etc. In addition, in supplement, an example of the context detection is given on the example of the social networks analysis.

The definition of probabilistic formal concepts and the proof of its properties is carried out under the assumption that a probabilistic measure is known. In practical tasks, we cannot assume this. Therefore, it is necessary to use a statistical criterion to detect MSCR on data. For that a Fisher 's independence exact criterion with some level of significance $\alpha$ was used. Then we can got a set $\mathcal{R}_\alpha$ of statistical approximations of MSCR detected on data with a confidence level $\alpha$ . They can cause contradictions in fixed points of probabilistic formal concepts. Therefore, it is necessary to introduce a criterion of approximations $\mathcal{R}_\alpha$ consistency in predictions on some data A. As a result we can discover a fixed points by consistently applying the prediction operator $\Upsilon(A)$ to data A, taking into account the consistency of predictions. The operator $\Upsilon(A)$ adds some new element to the set A if it is predicted by some approximation from $\mathcal{R}_\alpha$ based on data A and it is not contradict to the predictions of others by approximations in such a way that the consistency criterion strictly increases. Either the operator $\Upsilon(A)$ removes some element from A if there are approximations that predicting its negation and also in such a way that the consistency criterion again strictly increases. When the operator can no longer add a certain element to A, or remove an element in such a way that the operator strictly increases, we get a fixed point, which is denoted as $\Upsilon^{\infty}(A)$. A detailed description of this process is given in supplement of the work [12].

### 1.3 "Intelligent object" and "intelligent function"

The formalization of "Brain Principles Programming, (BPP)" described in [1] is based on the category theory and the concepts of "intellectual object" and "intellectual function" formulated in it. Here is an informal definition of "intellectual object" and "intellectual function" from [1]:

- "intellectual object", by which we mean any single integrity that we distinguish in this space – for example, when we see a table, signals from the optic nerve are processed by the brain, and the combination of individual lines is recognized as a table;
- the "intellectual function", which describes all possible operations in the system under consideration, is all that the psyche can do with an intellectual object. When we identify a table as an object, we can estimate its size or figure out how to use it;
- "essence" is the specific meaning of an object for the psyche. That is, the knowledge of what the table can be used for.

Formally, an "intelligent object" is defined as a mapping [1]:

- some data set (A);
- the observer as a reflection of the world after interacting with it (Ω);
- the relation of the world to the observer, which is a function of the internal state/ expectation (f) – an intellectual function $A \xrightarrow{f} \Omega$.

Here A is a data set that characterizes any unit wholeness;

Ω – distinguishing ability: "the relationship of an intellectual object with "me" does not yet mean any awareness, representation of this intellectual object in consciousness – it is enough for something to be somehow perceived and recognized to a degree sufficient for this "something" to be taken into account in the future in one way or another ... since the main task of thinking ... is to predict or produce a competitive future, then how the perceived object will eventually appear to us will depend on our mood, or, as phenomenologists would say, on our intentionality" [13]. More formally, "Some requirements will be imposed on the object Ω modeling the distinguishing ability, thus: first of all, it should be a partially ordered set, the elements of which correspond to "more" or "less" high values" [13].

Next, the waiting function $Exp_A : A \times A \to \Omega$ is defined (expectation) and the following explanations of this function are given: "at all levels of perception ... we are essentially dealing with a situation, with some expected state of affairs ... *Our psyche* *irresistibly tends to put the whole set of stimuli into a kind of understandable, clear and seemingly consistent picture of reality* (cursive 1 – E.V.) ... These representations of reality, in turn, are a specific filter-interpreter – any new stimuli, finding themselves, figuratively speaking, in the field of gravity of the corresponding system of representations, inevitably seem to change their trajectory – some are repelled (ignored), others, complementary, on the contrary, are attracted, others are modified (interpreted) in favor of the prevailing attitudes (cursive 2 – E.V.) ... As a result, with respect to any element x that is part of the intellectual object A, it is meaningful to say how different it is, firstly, from itself in the sense of what we expect to see in its place, and, secondly, how appropriate it is in the situation in general, i.e. how close it is to the rest of the elements distinguished in the situation."

More formally [13], the expectation function $Exp_A : A \times A \to \Omega$ assigns to each pair of elements $x, y \in A$ a measure of their consistency (coherence) on our existential partially-ordered scale Ω. At the same time, the measure of the object's $x \in A$ consistency with itself $Exp_A(x, x)$ can be understood as a measure of the proximity of x to its essence ... and denoted as $Ess_A(x)$ (essence). If we consider the prototype of class objects as an "invariant" of class objects, then the fixed point of the operator $\Upsilon^\infty(X(y))$ obtained on the set of properties $X(y) = \{P_1 \& ... \& P_m\}$ of some element $y \in A$ will differ from the attributes $X(y)$ of the element itself exactly as a measure of the object's consistency with itself. Therefore, the operator $\Upsilon^\infty$ gives a certain measure of the proximity of the object to its essence (invariant).

The waiting function allows to define the intelligent object more fully [13]. By an intelligent object we will understand ... an object $\mathcal{A} := (A, Exp_A)$ that includes a set of data A and an expectation function $Exp_A : A \times A \to \Omega$ that essentially depends on the subject experience Ω and its internal state.

The structure of an intelligent object described in cursive 1 above actually means that an intelligent object is a context, represented by a probabilistic formal concept, in which an operator $\Upsilon^\infty$ having the same meaning – minimizing contradictions in a set of stimuli – generates the most consistent picture of reality from the input set of stimuli A, supplementing it with all relevant information. At the same time (see cursive 2), new stimuli either change their trajectory, or repel, or attract. All these effects, which are taken into account in the expectation function $Exp_A : A \times A \to \Omega$, are modeled by the interaction of probabilistic formal concepts of elements of the set A. A change in the trajectory is the features moving closer to the prototype, repulsion is the braking of features, which is described in the supplement of the work [12], and attraction is mutual support at a fixed point.

**Intelligent function.** The work of the intellectual function is not only to recreate the intellectual objects associated with the data elements A and the set A itself, but also to connect these intellectual objects with all other intellectual objects that exist in the psyche and are relevant to a given situation, for example, to a need or some task (goal). The result of the work of the intellectual function is the creation of a "heavy intellectual object" by "enlarging the knowledge we have, which we believe relates to some problem that interests us" [13]. Thus, an intellectual object "is, as it were, raised to the degree of the knowledge (intellectual objects) that we possess, and acquires an appropriate meaning for us" [13].

Raising of some intellectual object "to the degree" of knowledge is formally represented as a relation: "If (this) relation is thought of as a certain kind of directional connection, then it seems quite natural to designate intellectual objects with letters $\mathcal{A},\mathcal{B},\mathcal{C}$ ..., and the relations between them with arrows ... $\mathrm{r}:\mathcal{A}\to\mathcal{B}$ " [13]. In addition, this relation "must respect those differences and identifications that were posited by the expectation function $Exp_A : A\times A\to\Omega$ " and satisfy the following conditions:

$\forall a,b\in A(Ess_B(r(a))\le Ess_A(a))$ , $Exp_A(a,b)\le Exp_B(r(a),r(b))$ .

All arrows of the ratio r, showing the path of enrichment of the intellectual object, form a "cone". The limit of the enrichment diagram is the cone generated by the "heavy" object contained in all other cones.

The operator $\Upsilon^\infty(A)$ automatically forms a context generated by elements A of some intellectual object and probabilistic formal concepts of its elements, since for all causal relationships linking elements A with other knowledge available in the psyche, the predicted elements of the psyche will automatically be included in the context by these causal relationships.

# Part II. Brain Principles Programming formalization

## 2.1 The principle of the complexity generation

The principle of the complexity generation in [1, p. 217] is formulated as follows: "The brain works with a very limited amount of information from the surrounding reality coming to its sensors … As this initially scarce information is used, the brain, at all levels of its organization, repeatedly increases its volume, correlating the received introductory data with the data already existing in it ... The principle of complexity generation allows the brain, having received the smallest external signal, to reproduce knowledge (an intellectual object) with incomparably greater power in the human mind, enriching the model of this object with information, which is relevant for the brain within the framework of its tasks (its goals)." This complexity generation is performed by the intellectual function introduced earlier in [1, p. 214], which "seems to be raised to the degree of the knowledge (intellectual objects) that we possess, and acquires an appropriate meaning for us".

The formalization of the "intellectual object" and "intellectual function" by probabilistic formal concepts gives us the following models of complexity generation:

1. If we consider the features of digits (see fig. 2 in supplement of the work [12]) as features perceived by the primary visual cortex, and the set A as a set of perceived digits, then the set of probabilistic formal concepts that have been discovered for these digits generates a set of intellectual objects $\mathcal{A}_0 := (0,\mathrm{Exp}_0)$ , $\mathcal{A}_1 := (1,\mathrm{Exp}_1)$ , ..., $\mathcal{A}_9 := (9,\mathrm{Exp}_9)$ – "invariants" of these digits. They are examples of generated complexity based on the simplest features (see description in supplement of work [12]).

2. The formation of contexts as probabilistic formal concepts, which generates a typology of social network users (see description in the supplement of the work [12]).

3. In general, when a certain task or a certain goal is set, the generation of complexity by the corresponding intellectual function will consist in generation of the context according to the given initial conditions A by "raising them to the degree" of the knowledge that is directly related to them. Formally, this generation is a probabilistic formal concept generated by an operator $\Upsilon(A)^{\infty}$ under conditions A, using all the knowledge related to the task or goal, represented by a set of MSCR approximations $\mathcal{R}_{\alpha}$ .

## 2.2 The principle of the relationship

In psychology, this principle was originally called the Gestalt principle. The brain reacts not to a specific stimulus, but to what this stimulus becomes when corresponding with the information that is already contained in the brain [1]. "The evaluation of information arising in the brain ... is carried out exclusively through the act of correlating one information with another, and the brain itself reacts not to the object of reality as such, but to how it correlates with other information located in the brain" [1].

Gestalt psychology puts forward the principle of integrity as the main explanatory principle. "The integrity of perception is a property of perception, consisting in the fact that every object, and even more a spatial objective situation, is perceived as a ***stable systemic whole***, even if some parts of it cannot be observed at the moment (for example, the back of a thing)" (Wikipedia). The integrity of perception, which is formed in the process of perception of a "natural" concept or prototype of a class, as well as the context of a certain task, is formally expressed in a probabilistic formal concept by cyclic mutual prediction of features of the concept or elements of the context. Therefore, the probabilistic formal concepts form the very "stable system whole" that characterizes integrity.

Therefore, formally, the operator $\Upsilon(A)^{\infty}$ is just the "stable system whole" in which, not individual elements of A are perceived, but their inseparable relationship with the rest of the fixed point $\Upsilon(A)^{\infty}$ elements.

## 2.3 The principle of approximation to the essence

The principle of approximation described in [1] as follows: "... in reality, there are no absolutely identical objects, so the brain performs approximation, that is, ignores differences if it manages to assign one or another "essence" to an object by specific features. At the same time, "essence" means the functionality of an object – what meaning it has for the brain (what role it performs) within the framework of the tasks it solves (its goals) ... when a person is tired and wants to rest – in the forest, a stump can serve as a chair, since you can sit on it".

The formation of "entities" occurs in the context of tasks being solved. Every context clarifies and correlates the elements of the context with a system of mutually predicted properties by causal relationships. This leads to the formation of "essences" associated with the context. For example, a knife in different contexts: cooking, combat situations, office work and hiking conditions should have different properties arising from the context: for a kitchen knife, the relationship of width, weight and blade edge is important, for a combat knife – the ratio of tip, length, weight and width, for a stationery knife – small weight, length and safety, for a penknife – relative smallness of size. Therefore, "essences" of "kitchen knife", "combat knife", "stationery knife", "penknife" are formed, which automatically generate probabilistic formal concepts in appropriate situations, since features and fix-points of their interrelation are different.

The context of the task, goal or need to be solved will automatically force you to choose the most suitable objects with the appropriate "functionality". This functional, which has a certain meaning for the brain within the framework of the tasks and goals it solves, will in a certain way affect the totality of the properties of the object, which, mutually assuming each other, and hence automatically form the corresponding probabilistic formal concept corresponding to its functional "essence".

Therefore, "essence" is a probabilistic formal concept $\Upsilon(A)^{\infty}$ generated by such elements A – features of the objects used, which will be selected in accordance with the context of the task being solved or the goal being achieved.

### 2.4 The principle of locality-distribution

The principle of locality-distribution [1]: "All information entering the brain can be duplicated many times in it, and its copies are processed in parallel by different structures independently, and only then this information is integrated into a holistic image." The brain processes information about a certain object in several modalities at once and in parallel – visual, auditory, tactile, etc. In each of these modalities, a hierarchy of the simplest "natural" classes and concepts is formed, for example, in the visual cortex, on the basis of perceived sticks, images of numbers can be formed, as in the mentioned example and "secondary" features – lines, angles, circles, etc., in the auditory cortex phonemes, words, text, etc. The integration of the image modalities is carried out through the perception of the integrity of the object, which integrates and binds the perception of parts into a "stable system whole" by some probabilistic formal concept.

Therefore, formally, this principle is also represented by an operator $\Upsilon(A)^{\infty}$ generating probabilistic formal concepts, not only for individual elements of A, but also for elements coming in parallel from different modalities.

### 2.5 The principle of heaviness

The principle of heaviness [1]: "The number of neural connections included in the creation of the object model, the number of relationships between the elements of the continuum of intelligent objects, the amount of information introduced into the object (attributes of the entity), the number of ways to calculate information about the object and the combination of multi-channel (modality) information about it into a single whole, correlated with the relevance of the task (goal) of the system, determine the "heaviness" of the intellectual object. The "heaviness" of the intellectual object determines the decision of the system. So, for example, if a person is hungry, he will look for food that will satisfy his hunger, but if he begins to face immediate danger (for example, from a predator), then a defensive strategy will begin to prevail, and he will stop looking for food and begin to escape".

In 1911, the dominant principle was put forward by A.A. Ukhtomsky [17]. This principle has also been preserved in the theory of functional systems [2-3], as the principle of dominance of the leading functional system, which creates the most "heaviness" context in case of satisfaction of some need.

In general, when it comes to solving a certain task or achieving a certain goal, possible solutions are obtained by different ways of enriching the original intellectual object "problem statement" or goal and form the corresponding "cones" and the contexts generated by them, which in the case of functional systems, we denote by a set $\{\mathcal{C} := (C, Exp_C)\}$. The choice of the "heaviest" of them is determined by the choice of the most desirable "heavy" solution.

Therefore, formally, the principle of heaviness consists in choosing the most desirable "heavy" intellectual object generated by one of the contexts $\{\mathcal{C} := (C, Exp_C)\}$ that are generated by the operator $\Upsilon^{\infty}(M \cup C)$, depending on the initial statement of the task/goal $\mathcal{M} := (M, Exp_M)$ and the available experience $C$ of solution of such a tasks/goals.

## 2.6 Conclusions

Algorithms for detecting probabilistic formal concepts, class prototypes, "natural" concepts and contexts are practically confirmed in the works (see supplement). The model of functional systems has also shown its effectiveness [13-16]. The integration of these algorithms and models can be carried out by using rules without actions as in the contexts of functional systems, assuming that the necessary actions will be performed and controlled by the corresponding functional systems. The integrated algorithm accurately models the basic cognitive functions of humans and animals mentioned in the first parts of the article, so the scope of applicability can be wide.

This approach can be generalized to a tasks-driven approach to the general artificial intelligence, as planned in [1] by generalizing functional systems to tasks-driven systems [18-20]. This approach solves the AGI problem formulated in [1] as: "general intelligence in AGI recognizes the ability to achieve goals in a wide range of environments, taking into account limitations". Therefore, Brain Principles Programming, formulated in [1] as principles of brain programming, based on research in cognitive sciences, can be implemented as a task-based approach to AGI, which, on the one hand, is able to solve a fairly wide class of tasks, and, on the other hand, corresponds fairly accurately to models of cognitive processes.

# Supplement

## A. Probabilistic formal concepts and their discovery

The formalization given below follows the works [21-24].

**Definition 1**. A *formal context* is a triple $\mathrm{K} = (\mathrm{G}, \mathrm{M}, \mathrm{I})$, where $G$ and $M$ are arbitrary sets of objects and attributes, and $I \subseteq G \times M$ is a binary relation expressing the attribute belonging to an object.

In a formal context, *derivative operators* play a key role – they connect subsets of objects and attributes of the context.

**Definition 2**. $A \subseteq G, B \subseteq M$, then:

$$A^{\uparrow} = \{m \in M \mid \forall g \in A, (g, m) \in I\}$$

$$B^{\downarrow} = \{g \in G \mid \forall m \in B, (g, m) \in I\}$$

**Definition 3**. A pair $(A, B)$ is a *formal concept* if $A^{\uparrow} = B$ and $B^{\downarrow} = A$.

Redefine the context within the logic, we will consider only finite contexts.

**Definition 4**. For a context $\mathrm{K} = (\mathrm{G}, \mathrm{M}, \mathrm{I})$ we define a *context signature* $\Omega_K$ that contains predicate symbols for each $\mathrm{m} \in \mathrm{M}$, $K \vDash m(x) \Leftrightarrow (x, m) \in I$.

**Definition 5**. For the signature $\Omega_K$, we define the following variant of first-order logic:

1. $X_K$ – a set of *variables*;

2. $\mathrm{At}_K$ – a set of *atomic formulas* (atoms) $\mathrm{m}(x)$, $m \in \Omega_K$, $x \in X_K$;

3. $\mathrm{L}_K$ – a set of *literals*, includes all atoms $\mathrm{m(t)}$ and their negations $\neg m(t)$;

4. $\Phi_K$ – a set of *formulas* defined inductively: literal is a formula, for any $\Phi, \Psi \in \Phi_K$ expressions $\Phi \wedge \Psi, \Phi \vee \Psi, \Phi \to \Psi, \neg\Phi$ – also formulas.

Let 's define the *conjunction* $\wedge L$ and *negation* $\neg L = \{\neg P \mid P \in L\}$ of a set of literals $L \subseteq \mathrm{L}_K$.

**Definition 6**. A single signature $\Omega_K$ element $\{\mathrm{g}\}$ forms a model $\mathrm{K}_g$ of this object. The truth of the formula $\phi$ on the model $\mathrm{K}_g$ is defined as $g \vDash \phi \Leftrightarrow K_g \vDash \phi$.

**Definition 7**. Let's define a *probability measure* $\mu$ on the set G in the sense of Kolmogorov. Then we can define a probability measure on a set of formulas as:

$\nu : \Phi_K \to [0,1], \nu(\phi) = \mu(\{g \mid g \vDash \phi\})$.

We assume that there are no non-essential objects in the context, such as $\mu(\{g\}) = 0, g \in G$.

**Definition 8**. Let $\{H_1, H_2, \ldots, H_k, C\} \in \mathrm{L}_K$, $C \notin \{H_1, H_2, \ldots H_k\}$, $k \geq 0$.

1. The *relationship* is $R = (H_1 \wedge H_2 \wedge \ldots \wedge H_k \to C)$;

The *premise* $R^{\leftarrow}$ of the relation R is a set of literals $\{\mathrm{H}_1, \mathrm{H}_2, \ldots, \mathrm{H}_k\}$;

The *conclusion* of the relationship is $R^{\to} = C$;

The *length* of the relationship is $|R^{\leftarrow}|$;

**Definition 9**. The *probability* $\eta$ of the ratio R is a value

$$\eta(R) = \nu(R^{\rightarrow} \mid R^{\leftarrow}) = \nu(R^{\leftarrow} \wedge R^{\rightarrow}) \big/ \nu(R^{\leftarrow}) .$$

If the denominator $\nu(R^{\leftarrow})$ of the ratio is 0, then the probability is undefined.

**Definition 10**. The relation $R_1$ is a *sub-relation* of the relation $R_2$, denoted as $R_1 \sqsubset R_2$ if $R_1^{\rightarrow} = R_2^{\rightarrow}$, $R_1^{\leftarrow} \subset R_2^{\leftarrow}$.

**Definition 11**. The relation $R_1$ *clarifies* the relation $R_2$, denote as $R_2 < R_1$ if $R_2 \sqsubset R_1$ and $\eta(R_1) > \eta(R_2)$.

**Definition 12**. The relation R is a *probabilistic causal relationship* if for each $\tilde{R}$ is fulfilled $(\tilde{R} \sqsubset R) \Rightarrow (\tilde{R} < R)$.

The definition of probabilistic causality given by Cartwright [20] with respect to some background can be formulated in these terms as follows. If the premise $R^{\leftarrow}$ of the relation R is a set of literals $\{\mathrm{H}_1, \mathrm{H}_2, ..., \mathrm{H}_k\}$ and we consider this set as a background, then each literal of the premise is a probabilistic reason for the conclusion $R^{\rightarrow}$ of the relation R with respect to this background, that is

$\nu(R^{\rightarrow} / R^{\leftarrow}) > \nu(R^{\rightarrow} / (R^{\leftarrow} \setminus H))$ for every $\mathrm{H} \in \{\mathrm{H}_1, \mathrm{H}_2, ..., \mathrm{H}_k\}$.

It is easy to see that this definition follows from definition 12.

**Definition 13**. The *strongest probabilistic causal relationship* is the relation R, for which there is no such probabilistic causal relationship $\tilde{R}$ that $(\tilde{R} > R)$.

**Definition 14**. *Semantic Probabilistic Inference* (SPI) of predictions of some literal C is a sequence of probabilistic causal relationships $R_0 < R_1 < R_2 ... < R_m$, $R_0^{\leftarrow} = \varnothing$, $R_m$ – the strongest probabilistic causal relationship and $R_0^{\rightarrow} = R_1^{\rightarrow} = R_2^{\rightarrow} ... = R_m^{\rightarrow} = C$.

**Definition 15**. The *tree of semantic probabilistic inference* Tree(C) of some literal C is the totality of all SPI predictions of the literal C.

**Definition 16**. The *most specific causal relation* for predicting some C is the strongest probabilistic causal relation of the tree Tree(C), which has the maximum conditional probability.

We denote by MSCR the set of all maximally specific causal relations for predicting some literal C. Under the system of causal relations, we will understand any subset $\mathcal{R} \subseteq$ MSCR.

**Definition 17**. The *prediction operator* for the system $\mathcal{R}$ is:

$$\Pi_{\mathcal{R}}(L) = L \cup \{C \mid \exists R \in \mathcal{R} : R^{\leftarrow} \subseteq L, R^{\rightarrow} = C\}.$$

**Definition 18**. The *closure* of the set of literals L is the smallest fixed point of the prediction operator containing L:

$$\Pi_{\mathcal{R}}^{\infty}(L) = \bigcup_{k \in \mathbb{N}} \Pi_{\mathcal{R}}^{k}(L).$$

A set of literals L is *consistent* if it does not simultaneously contain an atom C and its negation $\neg C$. A set of literals L is *compatible* if $\nu(\wedge L) \neq 0$.

**Theorem 1**. [24]. If L is compatible, then $\Pi_{\mathcal{R}}(L)$ compatible and consistent for any system $\mathcal{R}$.

**Definition 19**. A *probabilistic formal concept* on the context K is a pair (A, B) satisfying the following conditions:

$$\Pi_{\mathcal{R}}^{\infty}(B) = B,\ A = \bigcup_{\Pi_{\mathcal{R}}^{\infty}(C)=B} C^{\downarrow}.$$

The definition of the set A is based on the following theorem linking probabilistic and standard formal concepts on the context of K.

**Theorem 2**. [24]. Let $\mathrm{K} = (\mathrm{G}, \mathrm{M}, \mathrm{I})$ be a formal context, then:

1. If (A,B) is a formal concept on K, then there exists a probabilistic formal concept (S,T) on K such that $A \subseteq S,\ B \subseteq T$.

2. If (S,T) is a probabilistic formal concept on K, then there exists a family $\mathcal{C}$ of formal concepts on K such that

$$\forall (A,B) \in \mathcal{C}\ (\Pi_{\mathcal{R}}^{\infty}(B) = T),\ S = \bigcup_{(A,B)\in\mathcal{C}} A.$$

**The algorithm of statistical approximation of probabilistic formal concepts** [22,25]. In practical tasks, we cannot assume that the probability measure is known. Therefore, we need to use some statistical criterion to determine probabilistic inequalities in semantic probabilistic inference for MSCR discovery. To do this, we use the Fisher 's exact independence criterion with some level of significance $\alpha$. The resulting set $\mathcal{R}_{\alpha}$ of probabilistic maximally specific causal relationships obtained with the level of significance $\alpha$ can cause contradictions in fixed points of probabilistic formal concepts. Therefore, in order to approximate the operator $\Pi_{\mathcal{R}}(L)$, it is necessary to introduce an additional criterion for the consistency of the most specific causal relationships $\mathcal{R}_{\alpha}$ on the set L.

**Definition 20**. The causal relation $R \in \mathcal{R}_{\alpha}$ is *confirmed* on the set of literals $L$ if $R^{\leftarrow} \subset L$ and $R^{\rightarrow} \in L$. Then $R \in \mathrm{Sat}(L) \subseteq \mathcal{R}_{\alpha}$.

**Definition 21**. The causal relation $R \in \mathcal{R}_{\alpha}$ is *refuted* on the set of literals $L$ if $R^{\leftarrow} \subset L$ and $R^{\rightarrow} \in \neg L$. Then $R \in \mathrm{Fal}(L) \subseteq \mathcal{R}_{\alpha}$.

Now we can determine the criterion of maximum consistency of predictions based on the most specific causal relationships $\mathcal{R}_{\alpha}$ on a certain set of literals L.

**Definition 22**. The *criterion for maximum consistency of predictions* by the set of most specific causal relationships $\mathcal{R}_{\alpha}$ on the set of literals L is the value:

$$\mathrm{Int}(L) = \sum_{R\in\mathrm{Sat}(L)} \gamma(R) - \sum_{R\in\mathrm{Fal}(L)} \gamma(R).$$

The choice of causal relationship estimation $\gamma$ may depend on the specifics of the task. In our experiments we were guided by Shannon's considerations:

$$\gamma(R) = -\log(1 + \epsilon - \eta(R)),\ \epsilon > 0, \epsilon \ll 1.$$

Now we can approximate the operator $\Pi_{\mathcal{R}}(L)$ using this prediction consistency criterion.

**Definition 23.** We define the *operator* $\Upsilon(L)$ *of maximum consistency of predictions* for the most specific causal relationships $\mathcal{R}_\alpha$, which changes the set of literals $L$ by one element so as to strictly increase the criterion of consistency of predictions:

1. For all $G \in L_K \setminus L$, calculate the maximum increase in the criterion from the addition of G to $L$ : $\vartriangle^+ = \mathrm{Int}(L \cup \{G\}) - \mathrm{Int}(L)$ provided that there is in $\mathrm{Sat}(L)$ a pattern $R \in \mathrm{Sat}(L)$ such that $R^{\leftarrow} \subset L$ and $R^{\rightarrow} = G$.

2. For all $G \in L$, calculate the maximum increase in the criterion from the removal of G from L: $\vartriangle^- = \mathrm{Int}(L \setminus \{G\}) - \mathrm{Int}(L)$;

3. The operator $\Upsilon(L)$ adds the literal G to L if $\vartriangle^+ > 0$ and $\vartriangle^+ > \vartriangle^-$; the operator $\Upsilon(L)$ removes the literal G from L if $\vartriangle^- > 0$ and $\vartriangle^- > \vartriangle^+$. If $\vartriangle^- = \vartriangle^+$ and $\vartriangle^- > 0$, the operator $\Upsilon(L)$ removes the literal G;

4. If $\vartriangle^+ \le 0$ and $\vartriangle^- \le 0$, the operator $\Upsilon(L)$ returns L and, therefore, we have obtained a fixed point of maximally consistent predictions.

**Definition 24.** By *statistical approximation of probabilistic formal concepts* of context K for maximally specific causal relationships $\mathcal{R}_\alpha$, we mean the set of all fixed points that can be obtained as a result of applying the operator $\Upsilon(L)$ to some set of literals L representing some object $L = \{g\}^{\uparrow}$.

We show that in the limiting case when the set of regularities $\mathcal{R}_\alpha$ coincides with the system of causal relations $\mathcal{R} \subseteq \mathrm{MSCR}$, the fixed points of the prediction operator $\Pi_{\mathcal{R}}^{\infty}(B)$ and the operator $\Upsilon(L)$ of maximum consistency of predictions coincide. Therefore, statistical approximation of probabilistic formal concepts is a direct generalization of the original probabilistic formal concepts in the case of working with noisy data.

**Theorem 3.** Let $\mathcal{R}_\alpha = \mathcal{R} \subseteq \mathrm{MSCR}$. Then for any compatible set of letters L $\Upsilon^{\infty}(L) = \Pi_{\mathcal{R}}^{\infty}(L)$.

**Proof**: By virtue of Theorem 1 for $\mathcal{R}_\alpha = \mathcal{R}$ and L compatible we have $\mathrm{Fal}(L) = \varnothing$ at any stage of the application of the operator $\Upsilon(L)$. Then $\mathrm{Int}(L) = \sum_{R \in \mathrm{Sat}(L)} \gamma(R) > 0$ if $\mathrm{Sat}(L) \ne \varnothing$, since $\gamma(R) = -\log(1 + \epsilon - \eta(R)) = -\log(\epsilon) > 0$.

Then the inequality $\vartriangle^- = \mathrm{Int}(L \setminus \{G\}) - \mathrm{Int}(L) \le 0$ will always hold because $\mathrm{Sat}(L \setminus G) \subseteq \mathrm{Sat}(L)$. Therefore, according to definition 23, the operator $\Upsilon(L)$ will not remove the letters from $L$.

On the other hand, the operator $\Upsilon(L)$, in accordance with definition 23, will add new letters $G$ to $L$, provided that $\vartriangle^+ = \mathrm{Int}(L \cup \{G\}) - \mathrm{Int}(L) > 0$ and, this means that $\mathrm{Sat}(L) \subset \mathrm{Sat}(L \cup \{G\})$ and there is a rule $R \in \mathrm{Sat}(L \cup \{G\})$ such that $R^{\leftarrow} \subset L$ and $R^{\rightarrow} = G$. In fact, this means that the operator $\Upsilon(L)$ will add to $L$ one of the letters $G$ for which there is a rule $R^{\leftarrow} \subset L$ and $R^{\rightarrow} = G$.

Thus, in our case, the operator can be written as:

$$\Upsilon(L) = L \cup \frac{\arg\max \eta(G)}{\{G \mid \exists R \in \mathcal{R} : R^{\leftarrow} \subseteq L, R^{\rightarrow} = G,\}} .$$

Recall that the prediction operator has a similar form:

$$\Pi_{\mathcal{R}}(L) = L \cup \{C \mid \exists R \in \mathcal{R} : R^{\leftarrow} \subseteq L, R^{\rightarrow} = C\} .$$

The difference between the operators consists only in the sequence of adding letters, but both of them, with each application, add to the set those and only those literals $G$ for which there is a rule $\mathrm{R} \in \mathcal{R}$ such that $R^{\leftarrow} \subset L$ and $R^{\rightarrow} = G$. Since the order of adding letters does not affect the possibility of including other letters, the resulting fixed points $\Upsilon^{\infty}(L)$ and $\Pi^{\infty}_{\mathcal{R}}(L)$ coincide.

## B. Discovery of "natural" concepts and contexts

Here is an example of how the algorithm works by statistical approximation of probabilistic formal concepts for some context $\mathrm{K} = (\mathrm{G}, \mathrm{M}, \mathrm{I})$, where G is a set of encoded digits as shown in fig. 1, M is a set of features of digits (see fig. 1a) and I is the relation connecting digits and its features. For the experiment, a set of 360 shuffled digits was taken (12 digits of fig. 1 duplicated in 30 copies without specifying where which digit is). On this set, a set $\mathcal{R}_{\alpha}$ of 55089 probabilistic maximally specific causal relationships obtained with a significance level $\alpha = 0.01$ was found. Let $\mathrm{L}_K$ be the set of literals defined for all features, which we denote by predicates $L_k = \{P_1, ..., P_n\}$. Denote by $X(a) = \{P_1 \& ... \& P_m\}$ – the set of properties of some object $a$ given by predicates from $\mathrm{L}_K$.

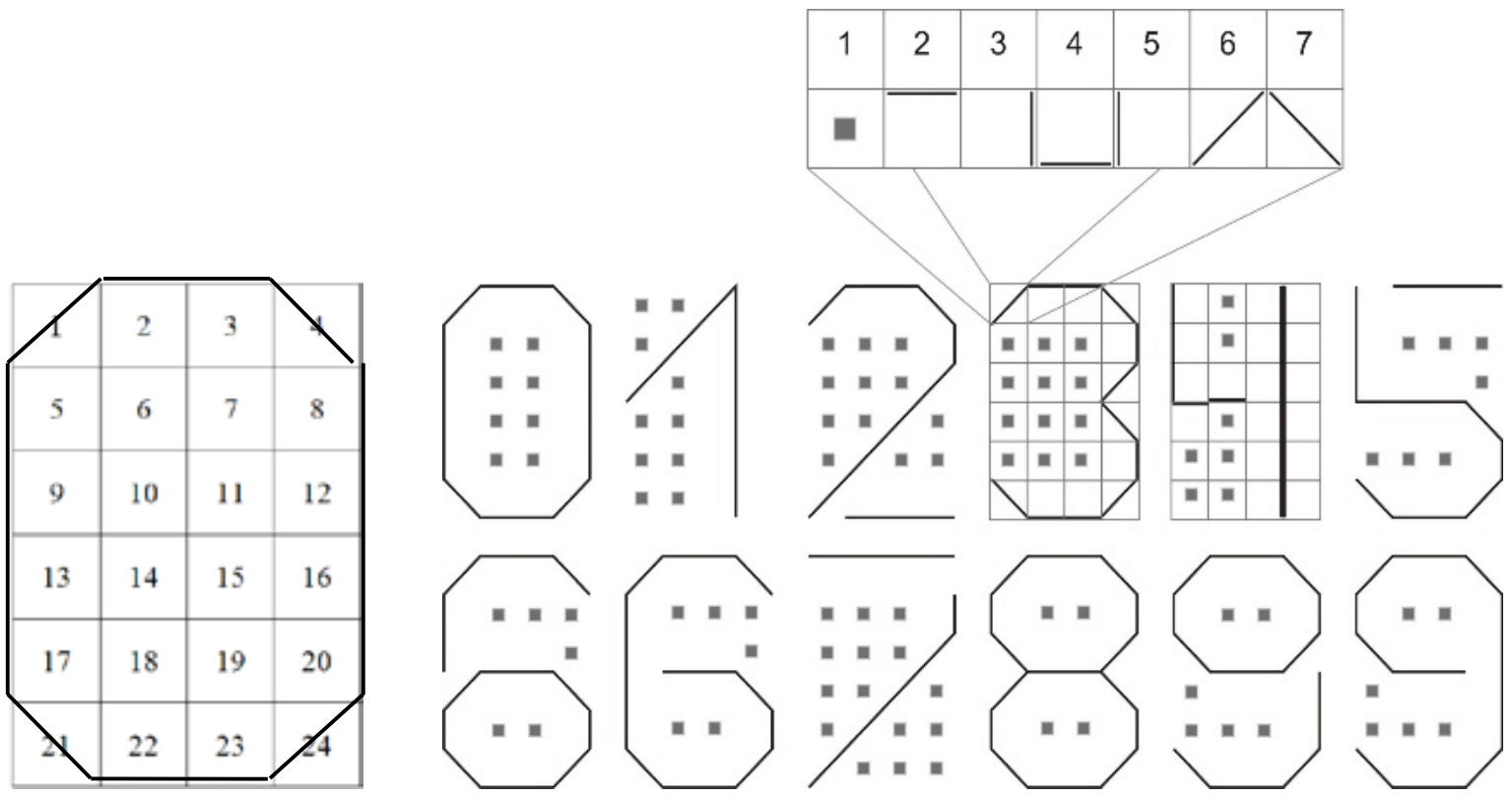

**Fig. 1**. Encoding of digits

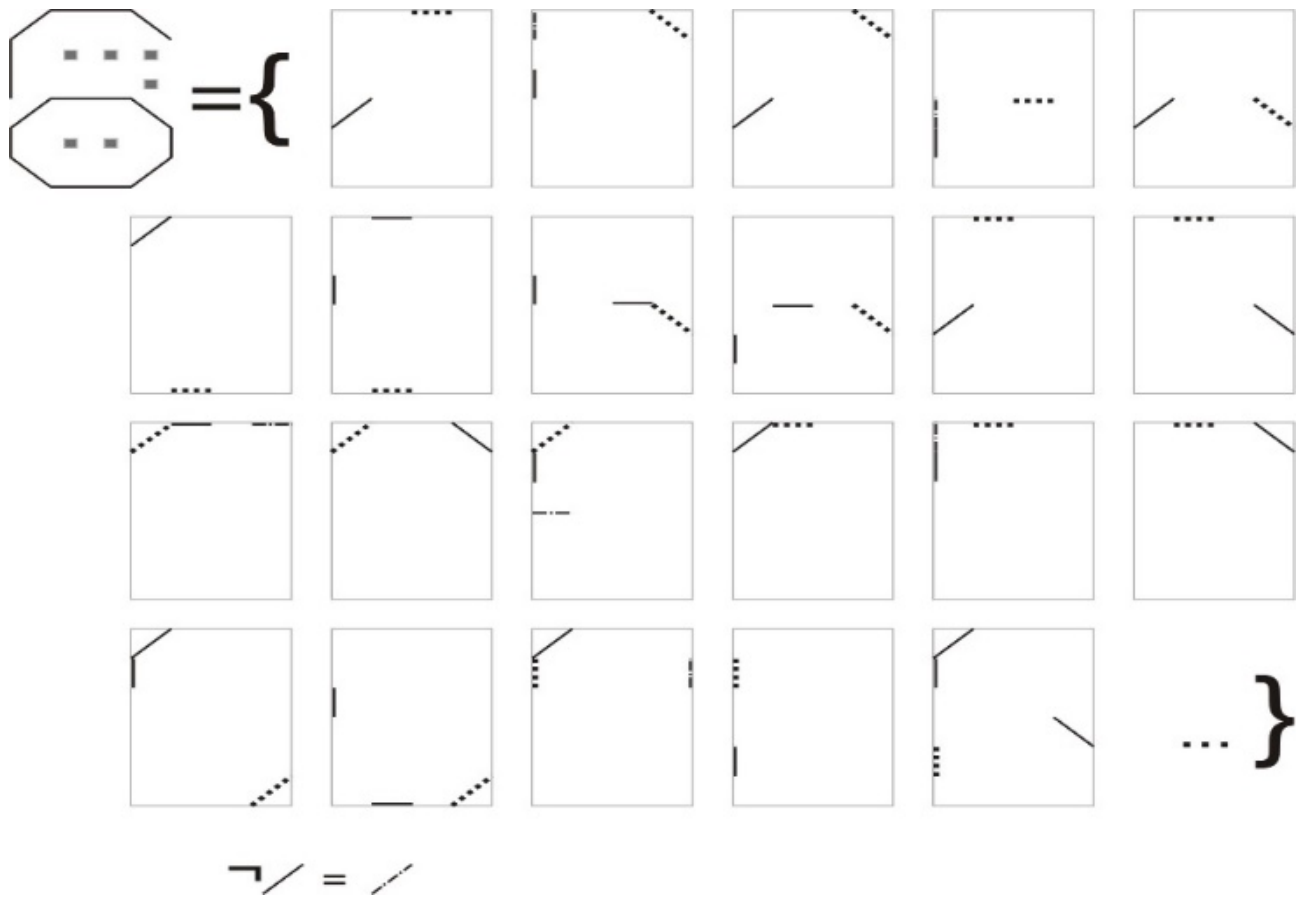

**Fig. 2.** Fixed point of the digit 6.

According to causal relationships $\mathcal{R}_\alpha$ , the operator $\Upsilon(L)$ found exactly 12 statistical approximations of probabilistic formal concepts that are fixed points of this operator and exactly correspond to 12 digits.

An example of a fixed point for the digit 6 is shown in fig. 2. Consider what this fixed point is. Let's numbering of the features of the digits as indicated in fig. 1 by numbers 1-24 and its values by signs 1-7 at the top of the fig. 1. The first causal relationship of the number 6 in fig. 2, represented in the first rectangle after the curly bracket, means that if there is a sign 6 in square 13 (denote it as 13-6), then there should be a sign 2 in square 3 (denote as (3-2)). The predicted feature is indicated by a dotted line. Let's write this causal relationship as (13-6 ⇒ 3-2). It is not difficult to verify that this relationship is carried out on all digits. The second causal relationship means that from the sign (9-5) and the negation of the sign 5 of the feature number first ¬(1-5) (the sign of the first feature should not be equal to 5), the sign (4-7) follows. Negation is indicated in the figure by a dotted line, as shown at the bottom of fig. 2. Then we get a causal relationship (9-5&¬(1-5) ⇒ 4-7). The next 3 patterns in the first row of the digits 6 will be respectively (13-6 ⇒ 4-7), (17-5&¬(13-5) ⇒ 4-7), (13-6 ⇒ 16-7).

In fig. 2 it can be seen that causal relationships and signs of the number 6 form a fixed point and mutually predict each other. Note that the causal relationships used in the fixed point are performed on all digits, but the fixed point itself distinguish only one digit. Therefore, the figures are distinguished not by the causal relationships themselves, but by their systemic interrelation. A fixed point forms a "prototype" according to Eleanor Rosch. The program does not know in advance which combinations of features are maximally correlated with each other.

It is important to note that causal relationships at a fixed point predict not only the presence of some other feature, but also the impossibility of the presence of any other feature at this fixed point. Thus, a fixed point is characterized not only by the presence of some signs in the corresponding squares, but also by the need for the absence of some other signs in some other squares, that is, it simulates the process of braking of some signs and corresponding prototypes of other classes.

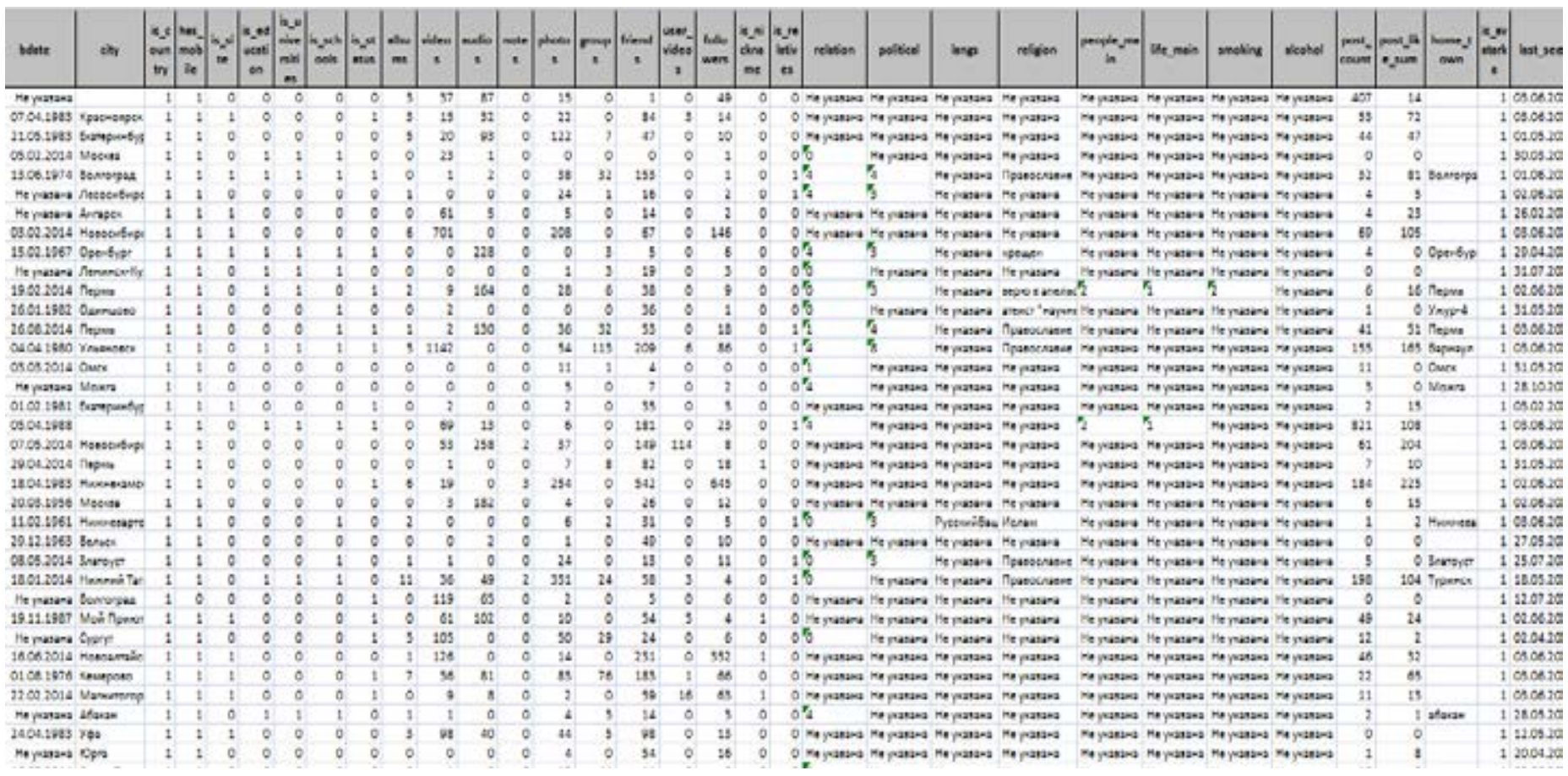

**Fig. 3**. User characteristics.

**The context formation on the example of the social networks analysis**. We will show how the statistical approximation of probabilistic formal concepts is applicable to the task of context formation. We study data on 2784 respondents (users) of the social network. We have data about 36 user's characteristics. Figure 3 shows the initial data. As a result of the statistical approximation algorithm, 21 fixed points of the operator $\Upsilon(L)$ were obtained, which in this case represent contexts or types of network users. The most demonstrative of them are the 8th and 11th contexts. The 8th context can be described as a married woman living in her hometown, for whom the main thing in life is family and children, values kindness and honesty in people, relatives are indicated in the profile, negative attitude to alcohol and smoking, a small number of photos, videos, audio. The 11th context can be described as an unmarried man, perhaps a teenager, who also values kindness and honesty in people, but the main thing for him in life is self-development, a compromise attitude to smoking and alcohol.

## C. Functional Systems Theory

Functional Systems Theory (TFS) is the leading theory of purposeful activity that describes the physiological mechanisms of goal-directed and purposeful behavior. By itself, thinking does not imply purposeful actions. We can plan the achievement of goals while lying on the couch. Therefore, we will divide purposeful behavior into two stages – the stage of planning actions and decision-making, which is carried out even before any actions, as the formation of a context that includes the "image of the desired future", and the stage of implementing purposeful behavior in accordance with the decision taken, together with the control of achieving intermediate and final results in accordance with the acceptor of actions results [2-3].

When a certain need arises, and, as a rule, a certain need is always dominates, then, firstly, there must be an element and an intellectual object corresponding to it in the set A, formed by motivational excitation, which activates the process of finding a solution to satisfy the need, and, secondly, a partially ordered set Ω, modeling our distinguishing ability by expectation function $Exp_A : A \times A \to \Omega$ of satisfying our need. It will evaluate elements A and their interaction from the point of view of satisfying the need and, accordingly, influence the construction of the intellectual object of the "image of the desired future". Therefore, the first stage can be considered as the context of a functional system for satisfying a need, formed by motivational excitation and the "image of a desired future". Detailed formalization of functional systems is given in [13-14].

Any functional system has the following architecture, the first stage of which we interpret as the context formation.

The **first stage** includes:

**Afferent synthesis**. Including the synthesis of motivational excitation, memory, situational and trigger afferentation:

- **Motivational excitation**. The goal setting in purposeful behavior is carried out by the need that has arisen, which is transformed into motivational excitation, which forms a basic intellectual object $\mathcal{M} := (M, Exp_M)$ that sets the distinguishing ability Ω, which will determine what is needed and what is not needed to satisfy the need.
- **Memory**. Motivational excitation "extracts from memory" all the sequences of actions that previously led to the goal achievement. Thus, the intellectual object $\mathcal{M} := (M, Exp_M)$ is "raised to a degree" – enriched by the experience of those cases ($\mathcal{C}$ - cases) that previously led to the satisfaction of this need. As a result, we get a set of enriched intellectual objects $\{\mathcal{C} := (C, Exp_C)\}$ corresponding to each case.
- **Situational and trigger afferentation**. Motivational excitation, taking into account the current situation and extracts from the memory only those experiences of achieving the goal that is possible in this situation.

**Decision-making**. The intellectual objects $\{\mathcal{C} := (C, Exp_C)\}$ of those cases that previously led to the satisfaction of the need are transferred to the decision-making block. According to the formalization [17-18], these methods include rules of the form:

$$P_1 \& ... \& P_n \& PG_1 \& ... \& PG_m \& A_1 \& ... \& A_k \Rightarrow PG_0 ,$$

Where $P_1 \& ... \& P_n$ is the condition of the situation required by this rule to achieve the goal, $PG_1 \& ... \& PG_m$ – are the sub-goals that need to be achieved to achieve the final goal $PG_0$ and $A_1 \& ... \& A_k$ – are the actions that, along with achieving the sub-goals, need to be performed to achieve the final goal $PG_0$. When some intellectual object $\mathcal{C} := (C, Exp_C)$ is "raised to the degree" of existing experience, taking into account the situation, it is enriched by such rules, but without taking into account actions, assuming that they will be fulfilled in the future. Therefore, the knowledge that the intellectual object is enriched with about the way to achieve the goal has a form that does not contain actions $P_1 \& ... \& P_n \& PG_1 \& ... \& PG_m \Rightarrow PG_0$. These rules will be included in the corresponding intellectual objects.

At the decision-making stage, only one of the ways to achieve the goal is selected, forming an action plan. Intellectual objects "raised to the degree" of the existing experience and having a certain way of achieving the goal, give "heavy" contexts corresponding to different "images of the desired future". Among these "heavy" contexts, the "heaviest" with the most desired "image of the desired future" is chosen. It is the resulting context of the first stage of the functional system.

**The acceptor of actions results**. The selected action plan corresponding to the selected "image of the desired future" also includes a sequence and hierarchy of all results that must be obtained to achieve the goal. The criteria for achieving these results, as a set of certain stimuli that must be obtained when they are achieved, form the acceptor of the actions results. These are certain "intellectual objects" with their own stimulation, expectation function and distinguishing ability, fixing the achievement of results.

**The second stage** consists in the implementation of the selected action plan in exact accordance with the "heaviest" context, together with the control of the achievement of intermediate and final results in accordance with the acceptor of actions results. If the goal is achieved and the need is satisfied, then this action plan is reinforced and recorded in memory.

In case of any deviation from the plan, an orientation-research reaction is activated, which revises the action plan and the "image of the desired future" and after that the formed context is revised.

In a series of experiments [15-17], the effectiveness of this scheme was confirmed.